\documentclass[twocolumn]{aastex631}
\usepackage{float}
\usepackage{bm}
\usepackage{ulem}
\usepackage{cancel}
\usepackage{soul}

\newcommand{\mat}[1]{\bm{\mathrm{#1}}}

\submitjournal{ApJS}

\begin{document}

\title{CFHT MegaCam Two Deep Fields Imaging Survey (2DFIS) II: Decoding the Lensing Profile of a "Rotating" Cluster with Deep CFHT Imaging}

\author{Yicheng Li}
\affiliation{Shanghai Key Lab for Astrophysics, Shanghai Normal University, Shanghai 200234, China}
\affiliation{Shanghai Astronomical Observatory, Chinese Academy of Sciences, Shanghai 200030, China}

\author{Liping Fu}
\affiliation{Shanghai Key Lab for Astrophysics, Shanghai Normal University, Shanghai 200234, China}

\author{Wentao Luo}
\affiliation{School of Aerospace Information and Technology, Hefei Institute of Technology, Hefei, Anhui 230031, China}
\affiliation{Department of Astronomy, School of Physical Sciences, University of Science and Technology of China, Hefei, Anhui 230026, China}

\correspondingauthor{Liping Fu}
\email{fuliping@shnu.edu.cn}

\correspondingauthor{Wentao Luo}
\email{wtluo@ustc.edu.cn}

\author{Binyang Liu}
\affiliation{Purple Mountain Observatory, Chinese Academy of Sciences, Nanjing, Jiangsu 210008, China}
% binyang_liu@alumni.brown.edu

\author{Wei Du}
\affiliation{Shanghai Key Lab for Astrophysics, Shanghai Normal University, Shanghai 200234, China}
% duwei@shnu.edu.cn

\author{Martin Kilbinger}
\affiliation{Universit\'e Paris-Saclay, Universit\'e Paris Cit\'e, CEA, CNRS, AIM, 91191, Gif-sur-Yvette, France}
% martin.kilbinger@cea.fr

\author{Calum Murray}
\affiliation{Universit\'e Paris-Saclay, Universit\'e Paris Cit\'e, CEA, CNRS, AIM, 91191, Gif-sur-Yvette, France}
% calum.murray@cea.fr

\author{Christopher J. Miller}
\affiliation{Department of Astronomy, University of Michigan, Ann Arbor, MI 48109, USA}
\affiliation{Department of Physics, University of Michigan, Ann Arbor, MI 48109, USA}
% christoq@umich.edu

\author{Ray Wang}
\affiliation{Department of Physics and Astronomy, Michigan State University, East Lansing, MI 48824, USA}
% wangru46@msu.edu

\author{David Turner}
\affiliation{Center for Space Sciences and Technology, University of Maryland, Baltimore County, Baltimore, MD 21250, USA}
\affiliation{High Energy Astrophysics Science Archive Research Center, Greenbelt, MD 20771, USA}
% djturner@umbc.edu

\author{Lance Miller}
\affiliation{Department of Physics, Oxford University, Keble Road, Oxford OX1 3RH, UK}
% Lance.Miller@physics.ox.ac.uk

\author{Dezi Liu}
\affiliation{South-Western Institute For Astronomy Research, Yunnan University, Kunming 650500, China}
% adzliu@ynu.edu.cn

\author{Mario Radovich}
\affiliation{INAF - Osservatorio Astronomico di Padova, via dell'Osservatorio 5, 35122 Padova, Italy}
% mario.radovich@inaf.it

\author{Jean-Paul Kneib}
\affiliation{\'{E}cole Polytechnique F\'ed\'erale de Lausanne, 1015 Lausanne, Switzerland}
% jean-paul.kneib@epfl.ch

\author{Huanyuan Shan}
\affiliation{Shanghai Astronomical Observatory, Chinese Academy of Sciences, Shanghai 200030, China}
% hyshan@shao.ac.cn

\author{Kaiwen Mai}
\affiliation{Shanghai Key Lab for Astrophysics, Shanghai Normal University, Shanghai 200234, China}

\author{Zicheng Wang}
\affiliation{Shanghai Key Lab for Astrophysics, Shanghai Normal University, Shanghai 200234, China}

\author{Haoran Zhao}
\affiliation{Shanghai Key Lab for Astrophysics, Shanghai Normal University, Shanghai 200234, China}

\begin{abstract}

We present a multi-wavelength analysis of the galaxy cluster RXCJ0110.0+1358 ($z=0.058$), a rotating cluster candidate, combining deep CFHT imaging, SDSS photometry, spectroscopic redshifts, and XMM-Newton X-ray observations. We find a clear difference between the optical galaxy distribution and the X-ray morphology: the catalog selected galaxies form two pronounced concentrations in the southeast (SE) and northwest (NW), whereas the X-ray emission exhibits a single, smoothly extended component centered near the SE brightest cluster galaxy. The combined galaxy sample exhibits significant spatially correlated kinematic substructure and a rotation like signal whose fitted projected axis approximately separates the two concentrations. Our weak lensing analysis identifies SE as the main detected mass concentration, spatially associated with the extended X-ray emission, and yields $\log(M_{200}/M_\odot)=14.04^{+0.24}_{-0.40}$ for this component. No significant weak lensing peak is detected at the NW optical center. Our results favor an interpretation in which the SE and NW galaxy populations, separated on the sky and differing in mean redshift, produce a rotation like signal when analyzed together.

\end{abstract}


\section{INTRODUCTION} \label{sect:intro}

% Paragraph 1: General introduction to galaxy clusters and their importance.
Galaxy clusters are the largest self-gravitating systems in the universe, consisting of tens to thousands of galaxies, the intracluster medium (ICM), and a dominant dark matter halo, with typical masses ranging from $10^{14}$ to $10^{15} M_{\odot}$. Representing the latest stage of hierarchical structure formation, galaxy clusters function as potent cosmological probes \citep{Kravtsov2012GCform, Allen2011}.

% Paragraph 2: The importance of mass and the unique role of weak gravitational lensing.
The mass of a galaxy cluster is a crucial probe for constraining cosmological models, necessitating precise measurement. Various techniques have been developed for mass estimation, including the Sunyaev-Zel'dovich (SZ) effect \citep{Bleem2015sz}, the temperature of the intracluster medium (ICM) from X-ray emission \citep{Ettori2013xraymass}, the dynamics of satellite member galaxies \citep{Heisler1985vir}, and gravitational lensing \citep{Kneib2011}. Notably, gravitational lensing is unique among these methods as it infers the total mass distribution directly, without relying on assumptions about the dynamical state or baryonic physics of the cluster. This is achieved by measuring the coherent distortion (shear) of background galaxy images, which arises from the deflection of light by the gravitational potential of the foreground cluster.

% Paragraph 3: The debate on cluster rotation, setting up the key scientific question.
An active area of research is the potential for galaxy clusters to exhibit coherent rotation. In the standard $\Lambda$CDM paradigm, halos are expected to acquire a net angular momentum from tidal torques exerted by the surrounding large scale structure during their formation \citep{Peebles1969}. However, observational claims of rotation are often based on kinematic signatures that can be ambiguous. These signatures include velocity gradients in the line-of-sight velocities of member galaxies (\citealt{HL07, tovmassian2015rot, MP17, bilton2019clrot}) or dipole patterns in kinetic Sunyaev-Zel'dovich (kSZ) effect maps (\citealt{Chluba2002rkz_clrot, Cooray2002rkz_halorot}). Such features may instead be transient phenomena caused by ongoing mergers, oscillating substructures, or simply the chance projection of unbound galaxy groups along the line-of-sight \citep{MP17}. Differentiating between a dynamically settled, rotating halo and a complex, merging system is therefore critical for understanding cluster assembly. Evidence from both simulations and observations remains contested, highlighting the need for multi-wavelength analysis (\citealt{Baldi2018rkz_clrot_sim, altamura2023simclrot, tang2025_clrot}).

% Paragraph 4: Introduction of THIS work, its motivation, and the specific target.
Distinguishing between these physical scenarios requires a direct probe of the underlying matter distribution, which motivates the use of weak gravitational lensing. A weak lensing mass map can reveal the fundamental morphology of the potential well. A disturbed, multi-peaked mass distribution would strongly indicate a merger, suggesting any kinematic signal is transient. Conversely, a singular, relatively symmetric mass peak located near the kinematic rotation center would be more consistent with a virialized halo where coherent rotation might be present. To this end, we initiated a program to map the mass of clusters identified as rotation candidates. From the SDSS DR7 \citep{SDSSDR72009} group catalog \citep{wang2014SDSSRASS, yang2007groupcat}, we selected 33 clusters with potential rotational or merging features, each comprising more than 50 member galaxies and possessing a mass exceeding $10^{14} M_\odot$. Specifically, these candidates are characterized by a sinusoidal variation in the mean redshift of member galaxies on opposite sides of the luminosity center as a function of the position angle of the dividing axis.

In this work, we present a detailed weak lensing analysis of one such candidate, RXCJ0110.0+1358, using deep observations from CFHT MegaCam \citep{Boulade2003}. This cluster is a particularly compelling candidate as it exhibits clear signs of a dynamically complex state. Optically, its galaxy distribution is bimodal, with two distinct luminosity centers in the southeast and northwest, resulting in a significant offset of approximately 10 $\mathrm{arcmin}$ between the luminosity weighted center and the Brightest Cluster Galaxy (BCG). By reconstructing its 2D mass map and comparing it with the galaxy kinematics and X-ray morphology, we investigate the physical origin of its previously reported rotation like signal. Our analysis identifies the primary weak lensing mass overdensity near the SE BCG, spatially associated with the extended X-ray emission. A single NFW fit to the SE tangential shear profile yields $\log(M_{200}/M_\odot)=14.04^{+0.24}_{-0.40}$, consistent within the quoted measurement uncertainties with the X-ray temperature based mass estimate. Together, these results favor an interpretation in which the SE and NW galaxy populations, separated on the sky and differing in mean redshift, produce a rotation like signal when analyzed together.

% Paragraph 5: Paper structure.
The remainder of this paper is structured as follows. Section~\ref{sect:data} describes the observational data, the XMM-Newton and CFHT data reduction procedures, and the X-ray surface brightness and spectral analyses. Section~\ref{sect:dyna} presents the galaxy kinematic analysis, including tests for substructure and rotation like signals. Section~\ref{sect:shape_measurement} describes the weak lensing analysis, including shear measurements, mass reconstruction, and halo mass modeling. Section~\ref{sect:wl_analysis} discusses the physical interpretation of our results, and Section~\ref{sec:sum} summarizes our main results.

% Paragraph 6: Adopted cosmology and constants.
Throughout this work, we adopt the AB magnitude system and a flat $\Lambda$CDM cosmology characterized by $\Omega_m = 0.28$, $\Omega_{\Lambda} = 0.72$, and $H_0 = 70~\mathrm{km\,s^{-1}\,Mpc^{-1}}$. At the cluster redshift of $z_\mathrm{cl} = 0.05828$, 1' corresponds to $\sim$ 68 kpc. The adopted cluster center is $\mathrm{ra}=01\text{h} 10\text{m} 03.17\text{s}$, $\mathrm{dec}=13^{\circ}58'41.38''$. The notation $M_{200}$ is utilized to denote the mass for which the mean density enclosed within a sphere of radius $R_{200}$ is 200 times the critical density of the universe.

\section{DATA} \label{sect:data}

RXCJ0110.0+1358, also known as Abell 152, was selected from the group catalog of \citet{wang2014SDSSRASS}, which combines ROSAT All-Sky Survey (RASS) X-ray information with the SDSS DR7 optical group catalog of \citet[][hereafter Y12]{Yang2012groupcat}. The optical groups were identified using model magnitudes and a modified halo based group finder \citep{yang2005groupfind}. The catalog assigns RXCJ0110.0+1358 a logarithmic halo mass of $\log(M/M_\odot)=14.43$.

We utilize the group catalog constructed by \cite{Yang2012groupcat}, which selects galaxies with extinction corrected $r$-band magnitudes $m_r<17.72$ and a redshift completeness qualification of $C>0.7$. Based on these criteria, the cluster comprises 71 member galaxies. Of these, 56 have spectroscopic redshifts from SDSS DR7, while the remaining members are assigned the redshift of their nearest neighbor in the Y12 catalog.
We adopt the BCG position as the cluster center. The BCG is identified as the brightest $r$-band galaxy with a spectroscopic redshift consistent with the cluster rest frame and is located at $\mathrm{RA}=01^{\mathrm h}10^{\mathrm m}03.17^{\mathrm s}$, $\mathrm{Dec}=13^\circ58^\prime41.38^{\prime\prime}$, with $z_{\mathrm{cl}}=0.05828$. In contrast, the luminosity weighted center cataloged in Y12, which derives luminosities from absolute magnitudes including $K$-corrections and evolutionary corrections, is positioned at $\mathrm{ra}=01\text{h}09\text{m}46.54\text{s}$, $\mathrm{dec}=14^{\circ}08'38.40''$, which is displaced by $\sim$11 arcmin from the BCG position with $z_\mathrm{cl}^{\prime}=0.05981$. This discrepancy arises due to the cluster's bimodal luminosity distribution, characterized by distinct southeastern and northwestern components. 

The system was detected in RASS and is included in the second release of the Meta-Catalogue of X-ray detected Clusters of galaxies (MCXC-II; \citealt{Sadibekova2024}). MCXC-II reports a mass estimate of $M_{500\mathrm{c}}=(4.26^{+0.73}_{-0.82})\times10^{13}\,M_\odot$, derived from the X-ray luminosity--mass scaling relation. The SE and NW regions were observed in two separate XMM-Newton pointings: ObsID 0503600701 for the SE field \citep{Arnold2009xray} and ObsID 0693221601 for the NW field \citep{Yershov2014xray}.

\subsection{X-ray Data} \label{subsec:xray}

We processed the XMM-Newton EPIC observations of both SE and NW fields, using SAS v22.1.0 and XMM-ESAS, following the ESAS cookbook\footnote{\url{https://heasarc.gsfc.nasa.gov/docs/xmm/esas/cookbook/xmm-esas.html}}. We screened soft proton flares using \texttt{espfilt} with $2\sigma$ thresholds for both observations and inspected the light curves. The final GTI durations, before live-time corrections, were 14.40, 15.42, and 5.74 ks for SE MOS1, MOS2, and pn, and 2.28, 2.70, and 1.32 ks for the corresponding NW detectors. Anomalous or inactive CCDs were excluded, and point sources were masked while retaining the central extended emission.

We constructed 0.5--2.0 keV images and radial profiles after subtracting the quiescent particle background (QPB), modeled residual soft proton emission, and pn OOT events; the photon sky background was retained. For each annulus, we combined the net counts and response scaled exposures from the three EPIC detectors to obtain the surface brightness in MOS2-equivalent count rate per unit solid angle. Profiles were measured from unsmoothed data in $30^{\prime\prime}$ wide annuli over $0$--$600^{\prime\prime}$, centered on the SE BCG and the NW optical center $\mathrm{ra=01h09m25.92s}$, $\mathrm{dec=14^{\circ}19'55.20''}$, respectively. We fitted the SE profile with a single $\beta$ model with a constant sky background by minimizing $\chi^2$. Detector specific predictions were PSF convolved and integrated over the annuli using the corresponding exposure maps and masks before combination. The fit gives $\beta=0.412\pm0.011$ and $r_c=(17.2\pm2.4)^{\prime\prime}$, with $\chi^2/\mathrm{dof}=18.46/16$. The quoted $1\sigma$ errors are derived from the local covariance matrix, with background templates, PSFs, and response factors held fixed.

\begin{figure*}
    \centering
    \includegraphics[width=1.75\columnwidth]{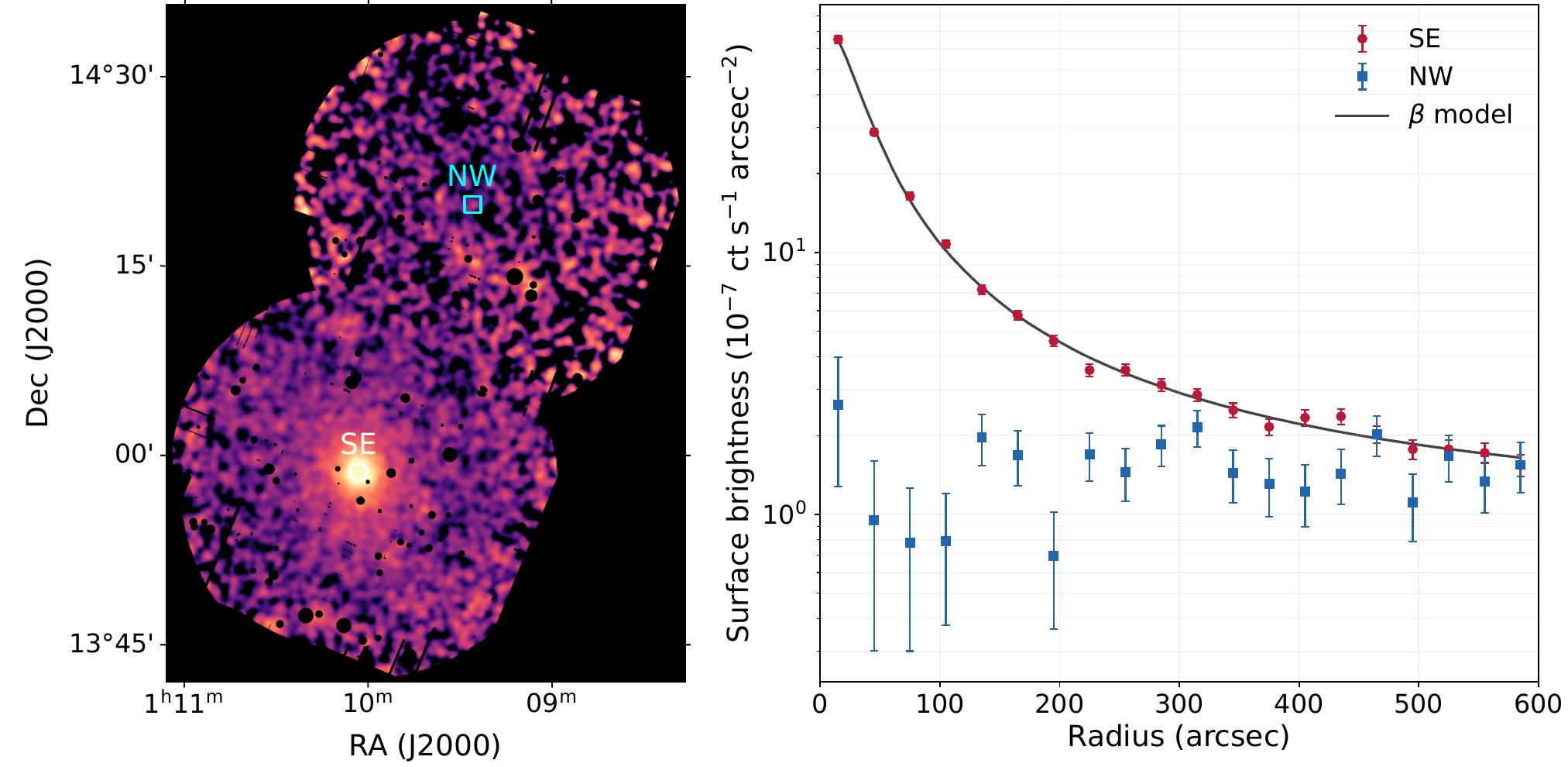}
    \caption{X-ray imaging and surface brightness profiles of RXCJ0110.0+1358. \emph{Left}: MOS2-equivalent 0.5--2.0 keV count rate map, with QPB, residual soft proton emission, and pn OOT events subtracted; the photon sky background is retained. Net-count and response-scaled exposure maps were separately smoothed with a Gaussian of $\sigma=15^{\prime\prime}$ before division. Markers indicate the SE BCG and NW optical center. \emph{Right}: Profiles centered on these positions, measured from unsmoothed data in $30^{\prime\prime}$-wide annuli. Red circles and blue squares denote SE and NW, respectively. The black curve is the annulus-averaged, PSF convolved $\beta$ model plus a constant sky background for SE. Error bars show $1\sigma$ statistical uncertainties at fixed background templates and response factors.}
    \label{fig:XMM_data}
\end{figure*}

The SE image and radial profile show centrally concentrated extended emission around the BCG (Figure~\ref{fig:XMM_data}). No comparable central enhancement is evident at the NW optical center, although the short exposure limits sensitivity to faint diffuse emission.

To maintain consistency with the XXL temperature definition adopted by \citet{umetsu2020mt} (hereafter U20), we measured the SE temperature within a core-included 300 kpc ($265.66^{\prime\prime}$) aperture centered on the BCG, using an absorbed single-temperature APEC model with the same fixed-abundance convention.\footnote{The metal abundance was fixed at $Z=0.3\,Z_\odot$, relative to the Anders \& Grevesse (1989) solar abundance table.} The redshift and Galactic absorption were fixed at $z=0.05828$ and $N_{\rm H}=3.57\times10^{20}\,\mathrm{cm^{-2}}$, respectively.

\begin{figure*}
    \centering
    \includegraphics[width=2\columnwidth]{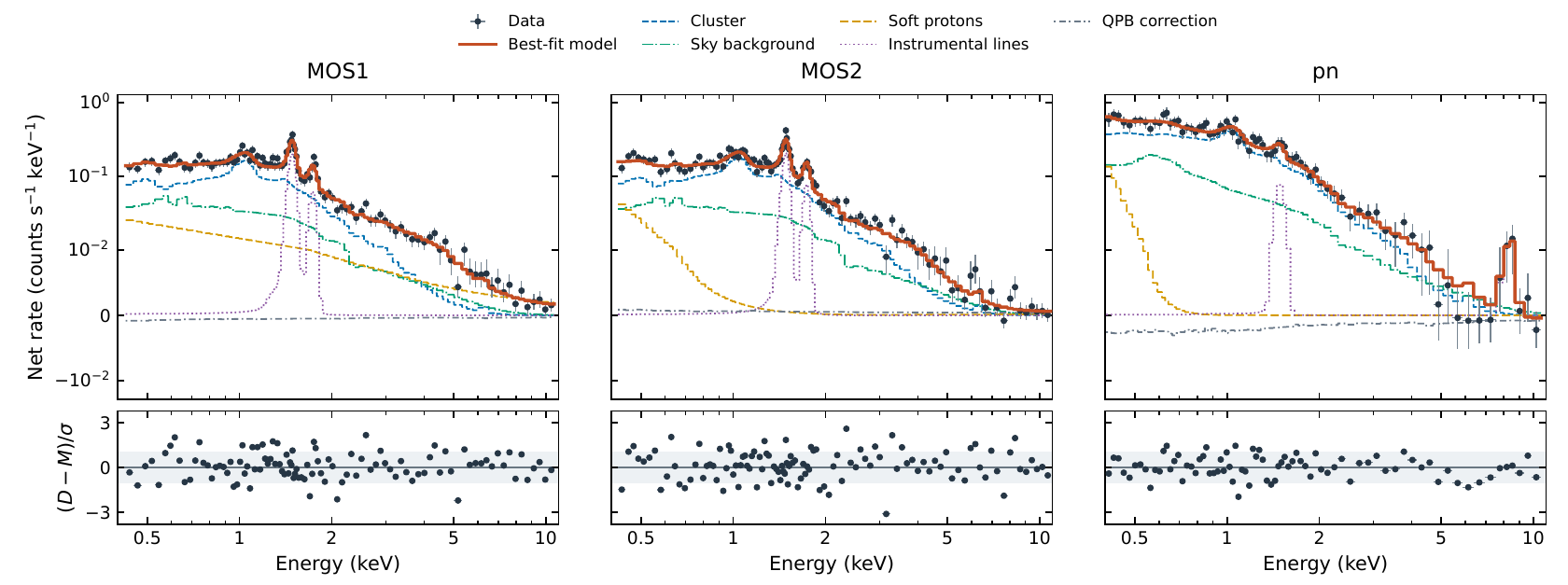}
    \caption{EPIC spectra within the core included 300 kpc aperture centered on the SE BCG. Columns show MOS1, MOS2, and pn. Upper panels display QPB-subtracted spectra and total best fitting models. Thin curves show individual model components, including the signed QPB correction $(a-1)B$, where $B$ is the subtracted ESAS template and $a$ its fitted scale factor. The vertical axes use a symmetric logarithmic scale, linear within $\pm0.01\,\mathrm{counts\,s^{-1}\,keV^{-1}}$. Lower panels show residuals $(D-M)/\sigma$, with shaded bands spanning $\pm1$. The EPIC fit covers 0.4--11.0 keV.}
    \label{fig:xray_spec}
\end{figure*}

Given the limited NW exposure, the SE source spectra were fitted jointly with spectra from a local $600^{\prime\prime}$--$810^{\prime\prime}$ annulus in the SE field and seven band RASS count rates from a $0.5^\circ$--$1.0^\circ$ annulus. The sky model included the Local Hot Bubble, Galactic halo, and unresolved cosmic X-ray background, with shared foreground surface brightnesses and region dependent CXB normalizations. To account for possible contamination of the local background annulus by thermal emission from the cluster, an additional absorbed APEC component with an independently fitted temperature and non-negative normalization in included in the local annulus. We subtracted the matched ESAS QPB spectra and included additive model corrections with the same spectral shapes, allowing the total QPB normalization of each source and local-annulus spectrum to vary independently within $\pm5\%$ of its nominal value. Residual soft proton emission and instrumental fluorescence were modeled separately. For both the pn source and local-annulus spectra, we used \texttt{pnback} to subtract exposure-normalized OOT spectra scaled by 0.063, extracted with the same screening, regions, and point source masks as the science spectra. We retained the statistical errors supplied by \texttt{pnback} for the OOT corrected spectra, with the associated QPB-background statistical uncertainties propagated in the fit. The EPIC spectra were grouped to at least 50 counts per bin and fitted using $\chi^2$ over 0.4--11.0 keV. The joint fit yields
\begin{equation}
kT_{\mathrm{300\,kpc}}=2.33^{+0.23}_{-0.20}\,\mathrm{keV},
\end{equation}
with $\chi^2=598.42$ for 663 degrees of freedom. The spectra and residuals are shown in Figure~\ref{fig:xray_spec}. The quoted $1\sigma$ intervals ($\Delta\chi^2=1$) were obtained by re-optimizing all other free parameters and are conditional on the adopted statistical error model, fixed QPB template shapes, and imposed normalization bounds. They exclude systematic uncertainties associated with these assumptions and other background and calibration effects. Fixing the QPB normalizations at their nominal ESAS values increases the best fitting temperature to $2.39$ keV. Replacing or supplementing the local background spectra with those from the NW observation changes the best fitting temperature by less than $0.01$ keV under either QPB prescription.

We estimate the SE halo mass using the weak lensing calibrated relation of \citet[][Equation~43]{umetsu2020mt},
\begin{equation}
E(z)M_{200\mathrm{c}}=(4.58\pm0.70)\times10^{13}\,M_\odot\left(\frac{kT_{\mathrm{300\,kpc}}}{1\,\mathrm{keV}}\right)^{3/2},
\label{eq:mt_scaling}
\end{equation}
where $E(z)=H(z)/H_0$ and $M_{200\mathrm{c}}$ is defined relative to 200 times the critical density. Using $E(0.05828)=1.02560$ and the unrounded best fitting temperature, we obtain
\begin{equation}
\log_{10}\left(M_{200\mathrm{c}}/M_\odot\right)=14.202^{+0.061}_{-0.058}.
\end{equation}
The quoted mass interval reflects only the conditional $1\sigma$ temperature uncertainty and excludes scaling relation normalization uncertainty, intrinsic scatter, mass calibration uncertainty, and spectral systematics. This estimate applies only to the SE X-ray halo.

\subsection{CFHT Imaging and Data Reduction} \label{subsec:img}
The galaxy cluster RXCJ0110.0+1358 was observed with the CFHT MegaCam on October 1, 2 and 4, 2020 (Proposal ID: 22BD07; PI W. Luo). MegaCam provides a $\sim 1~\mathrm{deg}^2$ field of view, which is large enough to cover the majority of the cluster's member galaxies in a single pointing. The total integration times of the $u$, $g$, $r$ and $i$ bands are 3840 s, 1680 s, 3680 s, and 3840 s, respectively, reaching a 5$\sigma$ point source limiting magnitude of approximately 26.0 mag in each band. A dithering strategy was applied to fill the CCD gaps. We used $r$-band images for the following weak lensing analysis.

A comprehensive reduction of the data was previously provided in \citet{liu2025overview} using the LSST Science Pipelines.
Here, we re-process the raw exposures with an independent workflow to ensure full compatibility with the shear measurement tools employed in this study.
This parallel reduction is complementary to \citet{liu2025overview} and is optimized for weak lensing shape measurements rather than general-purpose catalog production.
For each exposure, we perform cosmic ray rejection using the Python package \texttt{ccdproc} \citep{ccdproc} and sky background subtraction using \texttt{photutils} \citep{photutils}. Astrometric calibration is carried out using software \texttt{SCAMP} \citep{Bertin2006} with an external reference catalog from GAIA-DR2 \citep{Gaia2018}. The calibrated exposures were then coadded with \texttt{SWarp} \citep{Bertin2010} for the following processing.

Bad pixels and edges of images are masked with a flat field image. Masks of saturated stars are created by first selecting pixels close to the saturation threshold and then using a segmentation map that includes saturated pixels to mark saturated stars. Satellite traces are identified and masked by inspecting residual images created from subtracting individual exposures from the final coadd. Given the limited number of dithered exposures, bright star halos and scattered light artifacts were not fully suppressed by the coaddition process and were subsequently masked manually using \texttt{DS9}. The source catalog is then created using \texttt{SExtractor} \citep{Bertin1996}, 163,853 objects are detected in $r$-band coadded image when no selection is applied.

\section{DYNAMICS ANALYSIS} \label{sect:dyna}

\begin{figure}
    \centering
    \includegraphics[width=\columnwidth]{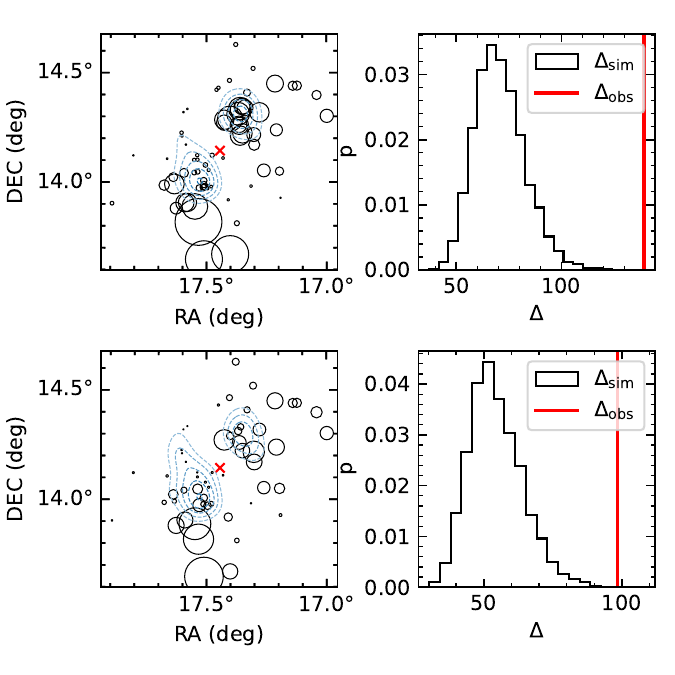}
    \caption{Results of the substructure test. The top and bottom rows correspond to Case 1 (all members) and Case 2 spectroscopic members only), respectively.
\emph{Left panels}: Spatial distribution of the $\delta$ statistic. The radius of each circle scales with $\mathrm{e}^{\delta_i}$. Blue contours trace the galaxy luminosity map, and the red cross marks the luminosity weighted center.
\emph{Right panels}: Distribution of the cumulative $\Delta_\mathrm{sim}$ statistic derived from Monte Carlo realizations. The vertical red line indicates the observed $\Delta$ value calculated from the data.}
    \label{fig:delta-test}
\end{figure}

\subsection{Tests of Substructure} \label{subsec:substructure}
As mentioned in Sect. \ref{sect:data}, there is a $\sim11 \,\text{arcmin}$ offset between the BCG and the luminosity weighted center, which suggests the presence of significant substructure. Therefore, following \cite{bilton2019clrot} (hereafter B19) and \cite{kang_deep_2024}, we applied the $\delta$-test (\citealt{Dressler1988_delta-test}) to investigate possible substructures, we approximate the line-of-sight velocity as $v = cz$, where $z$ is the redshift of galaxies from Y12 catalog. Given that the cluster catalog contains both spectroscopic and non-spectroscopic redshifts, we perform the analysis for two scenarios. For clarity, we define Case 1 as including all 71 identified members, and Case 2 as restricted to the 56 members with spectroscopic redshifts. The results for Case 2 are presented within parentheses below.

The $\delta-\text{test}$ quantifies the local deviation from the global mean velocity and velocity dispersion for each galaxy via the $\delta_i$ statistic, which is defined as:
\begin{equation}
    \delta_i^2 = \left(\frac{N_\mathrm{nn}+1}{\sigma^2_\mathrm{glob}}\right)[(\bar{v}_\mathrm{local}-\bar{v}_\mathrm{glob})^2 + (\sigma_\mathrm{local}-\sigma_\mathrm{glob})^2],
\end{equation}
where $\bar{v}$ and $\sigma$ are the mean velocity and velocity dispersion, respectively. The subscript "local" denotes the quantities computed from a subsample of $N_\mathrm{nn}$ nearest neighbors galaxies. We use $N_\mathrm{nn}=\sqrt{N_\mathrm{glob}}$ as suggested in B19, which is 9 for Case 1 (and 8 for Case 2) in this work. 
Using the line-of-sight velocities, we calculated the local and global statistics; the results are presented in the left panels of Figure \ref{fig:delta-test}. Each circle represents a galaxy, with its radius proportional to $\mathrm{e}^{\delta_i}$. Consequently, a clustering of large circles indicates the presence of potential substructures. As illustrated in the left panels of Figure \ref{fig:delta-test}, a clear concentration of galaxies with large $\delta_i$ values is observed in the northwestern part of the cluster, coinciding with the secondary luminosity peak. 

To quantify the statistical significance of these substructures, we calculate the cumulative statistic $\Delta$ defined as the sum of $\delta_i$ over all member galaxies. We perform 10,000 Monte Carlo realizations by randomly permuting the galaxy velocities while preserving their spatial coordinate. The resulting distribution of the simulated statistics, denoted as $\Delta_{\mathrm{sim}}$ is shown in the right panel of Figure \ref{fig:delta-test}. Notably, none of the 10,000 simulations produced a $\Delta_{\mathrm{sim}}$ value as large as the one observed ($\Delta_{\mathrm{obs}}$), implying a significance level of p-value $< 0.0001$. This result indicates significant spatially correlated kinematic substructure.

To further assess the one dimensional velocity distributions in both cases, we applied an Anderson--Darling test for Gaussianity and fitted Gaussian mixture models (\citealt{einasto2012multimodality}). Following the convention above, results for Case 2 are given in parentheses. The Anderson--Darling test yields $A^2=2.025$ $(1.014)$ with a $p$-value of $3.31\times10^{-5}$ $(1.05\times10^{-2})$. The evidence for non-Gaussianity is weaker in Case 2, where Gaussianity is rejected at the 5\% significance level but not at the 1\% level. We compared GMMs with one, two, and three Gaussian components, defining $\Delta\mathrm{BIC}_k=\mathrm{BIC}_k-\mathrm{BIC}_1$. The two- and three-component models yield $\Delta\mathrm{BIC}_2=-2.98$ $(+3.52)$ and $\Delta\mathrm{BIC}_3=-18.34$ $(+14.35)$, respectively. Thus, the three-component model has the lowest BIC in Case 1, whereas the single-Gaussian model has the lowest BIC in Case 2. Considering these assigned redshifts are not independent spectroscopic measurements, we place greater weight on Case 2 when interpreting the velocity distribution. The independent spectroscopic sample therefore retains evidence of non-Gaussianity, but the BIC comparison does not favor multiple Gaussian velocity components. This is compatible with the significant substructure detected by the $\Delta$-test, which incorporates spatial information in addition to velocities.

\subsection{Tests of Rotation} \label{subsec:rot}
\begin{figure*}
    \centering
    \includegraphics[width=\textwidth]{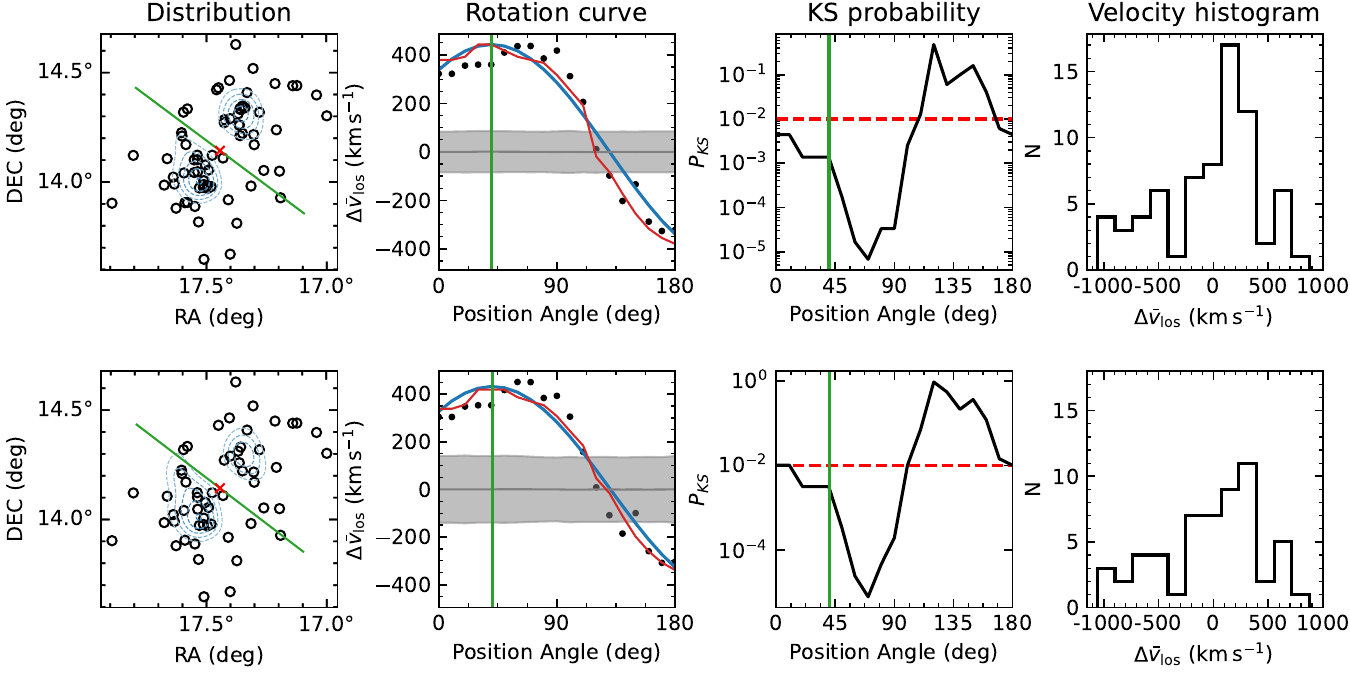}
    \caption{MP17 rotation analysis results for Case 1 (top row; all 71 members) and Case 2 (bottom row; 56 spectroscopic members).
\emph{Left}: Member galaxy distribution (black circle) with the best-fit rotation axis (green solid line).
\emph{Second column}: Rotation curves ($\Delta \bar v_\mathrm{los}$). Black dots: observed data; Blue curve: best-fit sinusoidal model; Red curve: ideal rotation profile; Gray curve/shading: random realizations with $1\sigma$ scatter.
\emph{Third column}: KS two sample test $p$-values. The red dashed line indicates $p=0.01$; the green line marks the best-fit axis angle.
\emph{Right}: Histograms of line-of-sight velocities.}
    \label{fig:rotation_signal}
\end{figure*}

Following the method proposed by \cite{MP17} (hereafter MP17), we searched for rotation signals by analyzing the spatial distribution of galaxy velocities. We calculated the rest frame line-of-sight velocity, $v_{\mathrm{los}}$, for each member galaxy as (\citealt{Danese1980}):
\begin{equation}
    v_{\mathrm{los}} = c \times \frac{z_{\mathrm{gal}} - z_{\mathrm{cl}}^\prime}{1 + z_{\mathrm{cl}}^\prime},
\end{equation}
where $z_{\mathrm{gal}}$ is the galaxy redshift and $z_{\mathrm{cl}}^\prime$ is the redshift of the cluster's rotation center.  To detect rotation, we defined an axis passing through this center on the sky plane, which divides the member galaxies into two subsamples. We then calculated the difference in the mean line-of-sight velocities between these two groups, defined as $\Delta \bar v_\mathrm{los} =  \bar v_{\mathrm{los},1} - \bar v_{\mathrm{los},2}$, where $\bar v_{\mathrm{los},1}$ and $\bar v_{\mathrm{los},2}$ are the mean velocities of the two groups. The velocity difference $\Delta \bar v_\mathrm{los}$ varies with the position angle ($\theta$). For a rotating galaxy cluster, this variation is expected to exhibit a sinusoidal pattern. The amplitude of this variation represents the projected rotational velocity, and the position angle yielding the maximum velocity difference corresponds to the projection of the rotation axis on the sky plane.

To further distinguish whether a cluster is rotating, MP17 introduced the ideal rotation curve and the random rotation curve. The ideal rotation curve represents a perfect rotation scenario. It is derived by assuming that the fitted rotation axis is the true axis. For this model, the line-of-sight velocity of each galaxy is set to a constant value: half of the rotation velocity. A positive or negative sign is then assigned to this value depending on which side of the axis the galaxy lies. 
The random curve represents the null hypothesis where no coherent rotation exists. It is generated by repeatedly and randomly shuffling the full set of galaxy velocities and reassigning them to the fixed galaxy positions on the sky. This process destroys any real kinematic correlation with position while preserving the overall velocity distribution.

Following the methodology of MP17, we quantify rotational significance using three statistical indicators: the reduced chi-square of the ideal model ($\chi_\mathrm{id}^2/\mathrm{dof}$), the ratio of the ideal-to-random chi squares ($\chi^2_{\mathrm{id}}/\chi^2_{\mathrm{rd}}$), and the Kolmogorov-Smirnov (KS) test probability ($P_\mathrm{KS}$).
The $\chi^2$ statistics are defined as:
\begin{equation}
    \chi^2_\mathrm{id} = \sum \frac{( {\Delta \bar v_{\theta} } - {\Delta\bar v_{\mathrm{id},\theta}} )^2}{\sigma_{\theta}^2 + \sigma_{\mathrm{id},\theta}^2},
\end{equation}
and
\begin{equation}
    \chi^2_\mathrm{rd} = \sum \frac{( {\Delta \bar v_{\theta}} - {\langle\Delta \bar v_{\mathrm{rd},\theta}\rangle} )^2}{\sigma_{\theta}^2 + \sigma_{\mathrm{rd},\theta}^2},
\end{equation}
where $\Delta \bar v_{\theta}$ is $\Delta \bar v_\mathrm{los}$ at position angle of $\theta$, $\Delta\bar v_{\mathrm{id},\theta}$ denotes the $\Delta\bar v_{\theta}$ from ideal rotation curve, and $\langle\Delta \bar v_{\mathrm{rd},\theta}\rangle$ represents the mean value derived from the ensemble of random realizations, $\sigma_{\theta}$ is the uncertainty at angle $\theta$, obtained using
\begin{equation}
    \sigma_{\theta}^2 = \left( \frac{\sigma_{v1}}{\sqrt{n_1}} \right)^2 +\left( \frac{\sigma_{v2}}{\sqrt{n_2}} \right)^2,
\end{equation}
the subscripts 1 and 2 denote two different sides of the rotation axis, $\sigma_v$ and $n$ are dispersion of the line-of-sight velocity and number of galaxies, respectively. According to MP17, a robust rotation detection requires satisfying three criteria simultaneously:
(1) The ideal rotation model provides a good fit to the data ($\chi_\mathrm{id}^2/\mathrm{dof}\leq1$); (2) The rotation signal is significantly distinguishable from random noise ($\chi^2_{\mathrm{id}}/\chi^2_{\mathrm{rd}}<0.2$); and (3) The velocity distributions of the two subsamples are statistically distinct ($P_\mathrm{KS}<0.01$).

We applied the MP17 methodology to investigate the rotation signal of RXCJ0110.0+1358. We acknowledge that MP17 suggests excluding substructures to prevent false positives. Excluding these high-$\delta$ members would effectively remove the primary signal of interest. Therefore, we apply the rotation analysis to the full sample to test whether this global kinematic structure aligns with an ordered rotation model.

Given the significant kinematic substructure signals detected in Sect. \ref{subsec:substructure} (and the large offset of the BCG), we adopted the luminosity weighted center as the reference center for rotation, since a rotation signal is often expected to be symmetric around the center of mass of the entire system. Figure \ref{fig:rotation_signal} illustrates the results for both cases. 
The differential velocity curve, $\Delta \bar v_\mathrm{los}(\theta)$, exhibits a sinusoidal variation with a maximum amplitude of $\sim443 \;(\sim432) \; \mathrm{km}\,\mathrm{s}^{-1}$ at a position angle of 40.17 (40.79) degrees. For the rotational scenario, we obtain a reduced chi-square of $\chi^2_{\mathrm{id}}/\mathrm{dof} = 0.230 (0.179)$, satisfying the first criterion. the derived rotation signal shows significant differences from the random expectation. The calculated ratio is $\chi^2_{\mathrm{id}}/\chi^2_{\mathrm{rd}}=0.026 (0.027)$, which falls well below the detection threshold of 0.2. Finally, we performed a two sample Kolmogorov-Smirnov (KS) test to compare the velocity distributions of the galaxy subsamples on either side of the rotation axis. This test yields a p-value of $P_\mathrm{KS} \approx 1\times10^{-3}(3\times10^{-3})$, which satisfies the significance threshold of $P_\mathrm{KS}<0.01$. 

The best-fit projected rotation axis approximately separates the SE and NW galaxy concentrations identified in Sect.\ref{subsec:substructure}.Together with the significant spatially correlated kinematic substructure, this alignment favors interpreting the rotation like signal as arising from the combined analysis of the SE and NW galaxy populations with different mean redshifts. Although the main detected mass and extended X-ray emission are localized in SE, the limited weak lensing sensitivity and short NW X-ray exposure do not exclude a low mass group or infalling structure at NW (see Sect.~\ref{subsec:xray} and Sect.~\ref{subsec:mass_recon}). The rotation like signal of the combined sample therefore does not provide secure evidence for coherent rotation of the SE main cluster.

\begin{figure}
    \centering
    \includegraphics[width=\columnwidth]{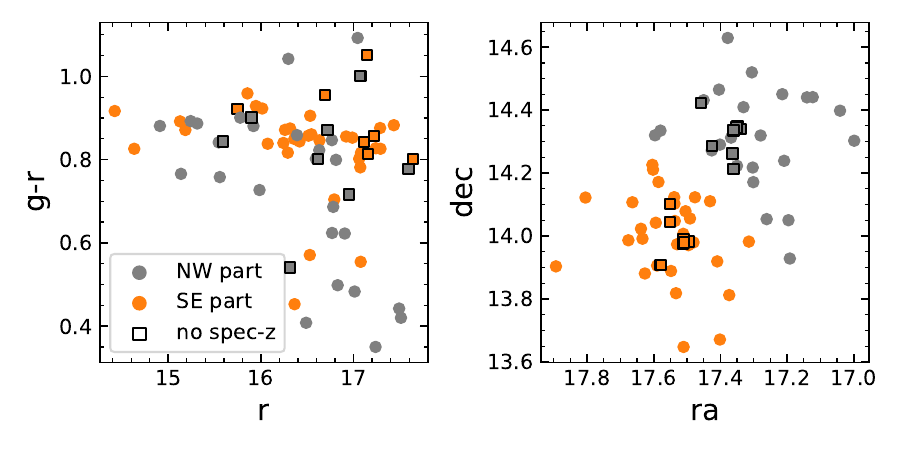}
    \caption{Properties of the member galaxies. \emph{Left panel}: Magnitude - Color diagram. \emph{Right panel}: Spatial distribution. In both panels, members associated with the northwestern and southeastern components are shown in orange and gray, respectively. Black hollow squares denote galaxies with redshifts estimated using the nearest neighbor method.}
    \label{fig:color}
\end{figure}
The northwestern and southeastern components, as separated by the best-fit rotation axis, also exhibit distinct properties in their galaxy populations and kinematics. Figure \ref{fig:color} presents the color-magnitude diagram using SDSS photometry, showing that while most member galaxies lie on the red sequence, a distinct population of blue galaxies is concentrated primarily within the northwestern component. Furthermore, treating the two components as spatially distinct groups reveals significant kinematic differences. The northwestern substructure is kinematically colder, with a velocity dispersion of $\sigma_{\text{NW}}=326\,(331)\,\text{km}\,\text{s}^{-1}$, compared to the hotter southeastern component, with $\sigma_{\text{SE}}=548\,(548)\,\text{km}\,\text{s}^{-1}$.

\section{SHAPE MEASUREMENT} \label{sect:shape_measurement}

\subsection{Basic weak lensing Theory} \label{subsect:shape_measurement_basic}

The deflection angle of light from background sources depends on the mass of the lens and the distances between the observer, lens, and background sources, which gives gravitational lensing the ability to infer the mass distribution of astrophysical objects. Weak lensing is widely used in studies of galaxy clusters; see \cite{umetsu2020clwl} and \cite{meneghetti2021WLintro} for  reviews. 

Weak lensing can be measured through the distortion of images, which can be described with the Jacobian matrix $\mat A$:
\begin{equation}
    \mat A = 
    \left(
    \begin{array}{cc}
        1 - \kappa - \gamma_1 & -\gamma_2 \\
        -\gamma_2 & 1 - \kappa + \gamma_1
    \end{array}
    \right),
\end{equation}
where $\gamma_1$ and $\gamma_2$ are the two components of shear, which stretch the image along two axes with an angle of $45^\circ$, while $\kappa$ is the convergence, which describes its isotropic magnification. The convergence $\kappa$ is related to the projected mass overdensity $\Sigma(\theta)$, which can be expressed as
\begin{equation}
    \kappa(\theta) = \frac{\Sigma(\theta)}{\Sigma_{cr}}.
\end{equation}
The critical surface density $\Sigma_{cr}$ is in the form of
\begin{equation}
    \Sigma_{\mathrm{cr}} = \frac{c^2}{4\pi G}\frac{D_\mathrm{s}}{D_\mathrm{l} D_{\mathrm{ls}}},
\end{equation}
where c and G are the speed of light and gravitational constant, $D_\mathrm{s}$, $D_\mathrm{l}$ and $D_\mathrm{ls}$ are the angular diameter distance between source and observer, lens and observer, lens and source, respectively. In practice, in the absence of information on the intrinsic size or magnitude of source galaxies, we can only observe the reduced shear $g = \gamma/(1 - \kappa)$.

In the weak lensing regime, the observed ellipticity $e_\mathrm{obs}$ is the complex sum of the intrinsic ellipticity of a source $e_\mathrm{s}$ and the reduced shear $g$ from lensing.  The reduced shear of a single galaxy is much smaller than its intrinsic ellipticity, which makes it difficult to measure for an individual galaxy. We can assume that the intrinsic ellipticities are randomly distributed, therefore their expectation value is $\langle e_\mathrm{s} \rangle = 0$. By stacking a large number of galaxies, the reduced shear can be estimated from the averaged observed ellipticities. 

\subsection{PSF Star Selection} \label{subsect:PSF_star}

Astronomical images suffer from the point spread function (PSF) caused by the optical system and the atmosphere if the telescope is ground based (see \citealt{liaudat2023PSF} for a review). An uncorrected PSF will bias shape measurements, a critical issue given that weak lensing distortions are typically at the one-percent level. The modeling and correction of the PSF is a crucial step before shape measurement.

Stars are ideal objects for PSF modeling, since they are natural point sources and are not distorted by the lensing effect. The PSF varies over time; therefore, PSF stars should be extracted on single exposure images.
We first run SExtractor for single exposure images, removing the artifacts by setting the DETECT\_MINAREA = 10, which corresponds to a radius of $\sim 2$ pixels. A pre-selection of FLAGS $\leq$ 4 is applied to the catalog of each exposure to exclude sources with bad results. The saturated stars are then removed according to the distribution of MU\_MAX. PSF star candidates are then selected from the stellar locus in the MU\_MAX $-$ MAG\_AUTO diagram, where unsaturated stars form a tight, well defined sequence.

We applied a magnitude cut of 22.5 to exclude those objects that might be contamination from galaxies. Finally, we select those stars with FLAGS == 0 to exclude sources with bad pixels or blended with other objects. For each exposure, about 2,500 PSF stars are selected for PSF modeling.

\subsection{Lensfit PSF Modeling and Shape Measurement} \label{subsect:lensfit}

% intro of lensfit
\texttt{lensfit} \citep{miller2013lensfit} was first used in CFHTLenS \citep{heymans2012CFHTLenS} for shape measurement, and it proved to be a powerful tool for weak lensing analysis. Given that our data were also taken with the CFHT MegaCam, we adopt the well-tested \texttt{lensfit} pipeline for both PSF modeling and shape measurement.

\begin{figure}
    \centering
    \includegraphics[width=\columnwidth]{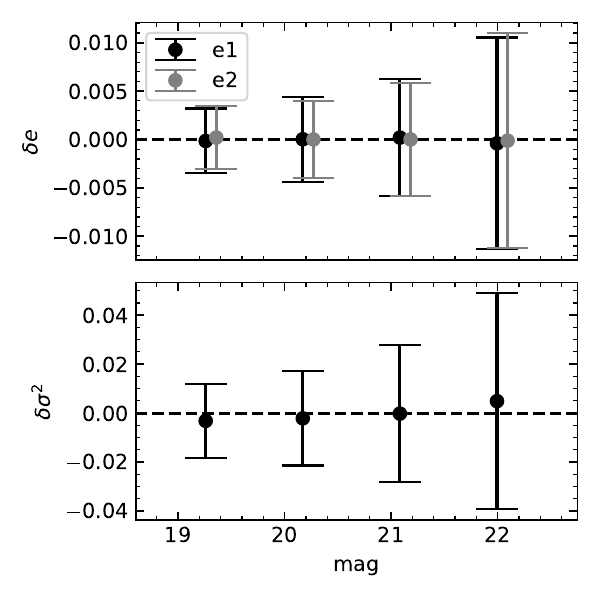}
    \caption{Residuals of ellipticity and size between PSF model and PSF star.}
    \label{fig:psf_result}
\end{figure}

\texttt{lensfit} models the PSF on a pixel grid. To account for spatial variations across the field of view, the value of each pixel in this model is described by a polynomial function of the position on the detector mosaic. Following \cite{fu_weak_2018}, we configure this built-in model by setting its key parameters: the polynomial order for the global, field-of-view variation is set to four, while the order for chip dependent variations on individual CCDs is set to one. Figure \ref{fig:psf_result} shows the fitted PSF model and the residual between the PSF model and stars of 4 exposures. The average ellipticity before correction is $\langle e_1 \rangle = 2.17\times10^{-3}$ and $\langle e_2 \rangle = 1.86\times10^{-3}$, after correction, the mean residual ellipticities are reduced to $\langle e_1 \rangle = -4.71 \times 10^{-4}$ and $\langle e_2 \rangle = 1.17 \times 10^{-5}$, showing a significant decrease, as expected. 

After PSF modeling, shear measurements are carried out with a likelihood based method, the ellipticity is then given by fitting the galaxy postage stamps with a PSF convolved two-component (bulge + disk) model. The detailed algorithm of \texttt{lensfit} are described in \citealt{miller_bayesian_2007}, \citealt{kitching_bayesian_2008} and \citealt{miller2013lensfit}.

\subsection{Source Selection} \label{subsect:sourcc_selection}

The galaxy cluster has a low redshift of 0.05828, according to the photo-$z$ catalog from CFHTLS (\citealt{Hildebrandt2012CFHTLS}), only 1.21\% of galaxies have $z_\mathrm{phot}<0.058$, which means that the lensing signal is dominated by the background sources. Furthermore, with photometric data in only four bands, the resulting photo-$z$ accuracy is insufficient to reliably separate foreground member galaxies from the background source population. Therefore, we only applied a magnitude cut of $21 < m_\mathrm{r} < 26$ on the selection of background sources.

% selection by lensfit
In addition to the magnitude cut, \texttt{lensfit} has a strict selection of background sources before and during the fitting process. Objects that are too large for the postage stamp, are heavily blended, or fail the model fitting procedure are assigned a weight of zero and are thus excluded from the final shear catalog. For the other galaxies, a weight and flag are assigned, the weight is based on shape noise and ellipticity-measurement noise. To further increase the reliability, we select the galaxies observed in 4 exposures. Additionally, we applied a size cut of the final shear catalog, excluding galaxies with $r_\mathrm{ab} < 0.3 \, r_\mathrm{PSF}$, where $r_\mathrm{ab}$ indicates the geometric mean of the major and minor-axis scalelength, and $r_\mathrm{PSF}$ is the $r_\mathrm{ab}$ of PSF model at the position of galaxy. The final catalog contains 53,006 source galaxies, corresponding to a number density of approximately 15 galaxies $\mathrm{arcmin}^{-2}$.

\subsection{Shear Calibration}
The measured shear can be biased by a variety of instrumental and astrophysical effects \citep{meneghetti2021WLintro}. Generally, the observed reduced shear, $g^{\text{obs}}$, is modeled as being linearly biased with respect to the true shear, $g^{\text{true}}$, by a multiplicative coefficient, $m_i$, and an additive coefficient, $c_i$.
\begin{equation} \label{eq:shear_bias}
    g_i^\mathrm{obs} = (1+m_i)g_i^\mathrm{true} + c_i.
\end{equation}

\begin{figure*}
    \centering
    \includegraphics[width=\linewidth]{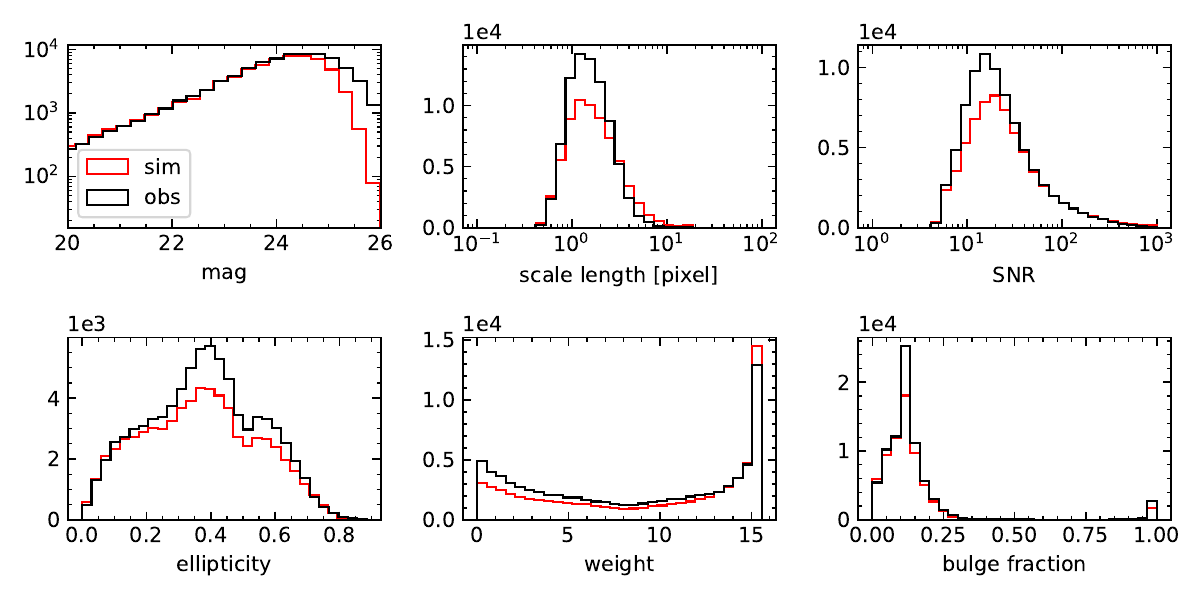}
    \caption{Comparison of observed data (black) and mock data (red) for distributions of galaxy properties. From left to right, top to bottom: magnitude, scale length, signal to noise ratio, ellipticity, \texttt{lensfit} weights and bulge fraction.}
    \label{fig:mock_img_check}
\end{figure*}

Following \cite{Liu2018biascal}, shear calibration is carried out with image simulation based on \texttt{Galsim} \citep{Rowe2015}. The mock galaxy catalogs are generated with the same coordinates and magnitude as the observed data, which ensures the distribution of galaxies in a more realistic way. Galaxies are generated with the same two components used in \cite{miller2013lensfit}, with a bulge-to-total flux ratio ($B/T$) which is a truncated Gaussian distribution with $\mu = 0.0$ and $\sigma = 0.1$. 10\% of galaxies are set as bulge dominanted galaxies with $B/T = 1.0$. The S\'ersic index distribution of disk galaxies is truncated Gaussian with $\mu = 0.72$ and $\sigma=\sqrt{0.15}$, ranging from 0.3-4.0, where indices exceeding 4.0 are set to 4.0. For bulge dominated galaxies, the S\'ersic index is set to 4.0. The scale length radius follows the distribution given in \cite{Kuijken2015},
\begin{equation}
    \log_e (r_\mathrm{d}) = -1.320 - 0.278(r-23).
\end{equation}
The intrinsic ellipticity of disk galaxies follows the distribution given in \cite{miller2013lensfit}
\begin{equation}
    p(e) = \frac{Ae(1-\exp((e-e_\mathrm{max})/a))}{(1+e) (e^2 + e_0^2)^{1/2}}, (e<e_\mathrm{max}),
\end{equation}
where we adopted  $a=0.2539$, $e_\mathrm{max} = 0.803$ and $e_0 = 0.0256$ from \cite{miller2013lensfit}. For bulge dominated galaxies, the distribution of ellipticity is
\begin{equation}
    p(e) \propto e \exp(-be-ce^2),
\end{equation}
with $b=2.368$ and $c=6.691$.

We applied the same WCS information from the header of the observations on the mock images to reproduce the dithering pattern of the observation. The mock images are then generated on the basis of the mock galaxy catalog. We apply a constant shear modulus of $|\gamma|=0.04$ along four directions, (+0.0283, +0.0283), (-0.0283, -0.0283), (-0.0370, +0.0153) and (+0.0153,-0.0370).To suppress shape noise, we employ the shape-pair cancellation technique by rotating the galaxy images by $90^\circ$. The total number of mock galaxies is $\sim1.4\times10^6$.

For each set of images, we applied the same PSF models fitted from the observation data. Noise is added according to sky noise, readout noise, and gain in headers. This noise, combined with random galaxy shapes, resulted in the non-detection of faint sources, reducing the final effective number of galaxies. Figure \ref{fig:mock_img_check} shows the comparison of the mock and obs different parameters given by the \texttt{lensfit} for galaxies with weight $>$ 0. 

\begin{figure}
    \centering
    \includegraphics[width=\columnwidth]{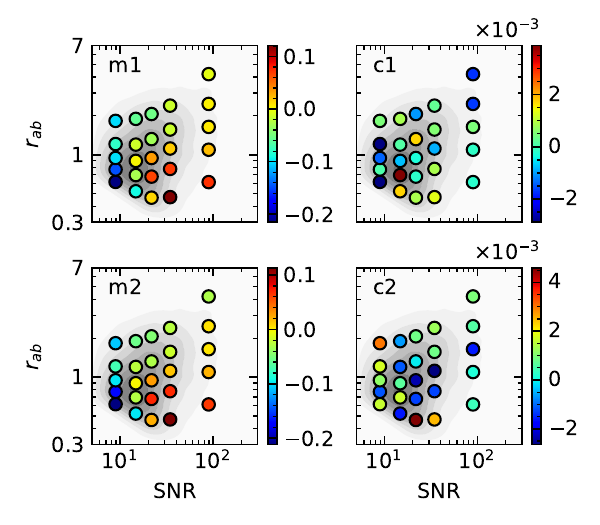}
    \caption{Shear bias coefficients on SNR-$r_\mathrm{ab}$ surface. \emph{Right panels} are the multiplicative bias, \emph{left panels} are the additive bias. Background filled contours indicate the number of galaxies.}
    \label{fig:bias_cal}
\end{figure}

We followed \cite{Liu2018biascal} and binned the galaxies on a $SNR-r_{ab}$ surface, where $r_{ab}$ is the geometric average of the major and minor axes in scale length. Considering that our mock sample size is relatively small, the galaxies are binned into $5\times 5$ bins, such that the number of galaxies in each bin is identical. For each bin, we fitted the input shear and measured shear with Eq. \ref{eq:shear_bias} to get the additive and multiplicative bias coefficients, the results are shown in Figure \ref{fig:bias_cal}. The shear of observed galaxies are then calibrated according to the bins they fall into.

\subsection{Systematic Tests}  \label{subsec:sys}
\begin{figure}
    \centering
    \includegraphics[width=\linewidth]{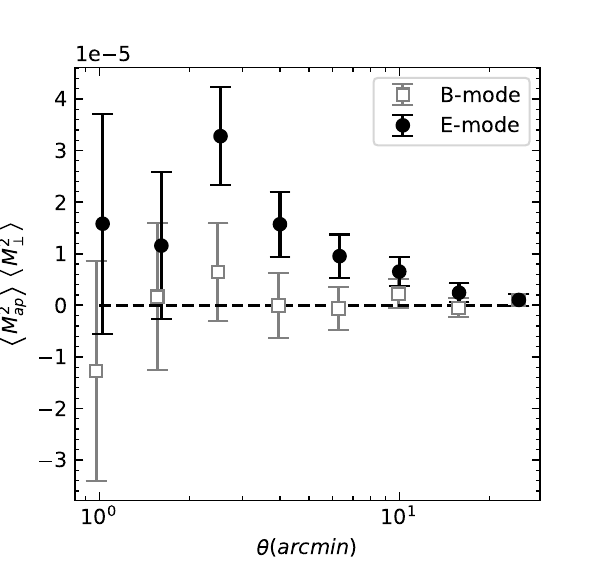}
    \caption{E-/ B-mode from aperture mass dispersion.}
    \label{fig:EBtest}
\end{figure}

Weak lensing is in general sensitive to systematic errors. In an ideal case, the gravitational shear field is curl-free and purely constitutes an E-mode. Consequently, the presence of a divergence-free B-mode serves as a diagnostic for residual systematics. Before mass map construction and mass estimation, we performed an E- and B-mode separation test using the aperture mass dispersion from the shear two-point correlation function (2PCF). It is well established that observable B-modes mainly originate from imperfections in image processing and data analysis pipelines \citep{kilbinger_cosmology_2015}, with intrinsic galaxy alignments producing subdominant contributions on small scales. Figure \ref{fig:EBtest} presents the measured results of the E / B mode decomposition. The B-mode is close to 0 at all scales, while a significant E-mode is observed at $2^\prime$ to $10^\prime$. 

\section{RESULTS and DISCUSSION} \label{sect:wl_analysis}

\subsection{Mass Reconstruction and Estimation} \label{subsec:mass_recon}

\begin{figure}
    \centering
    \includegraphics[width=\columnwidth]{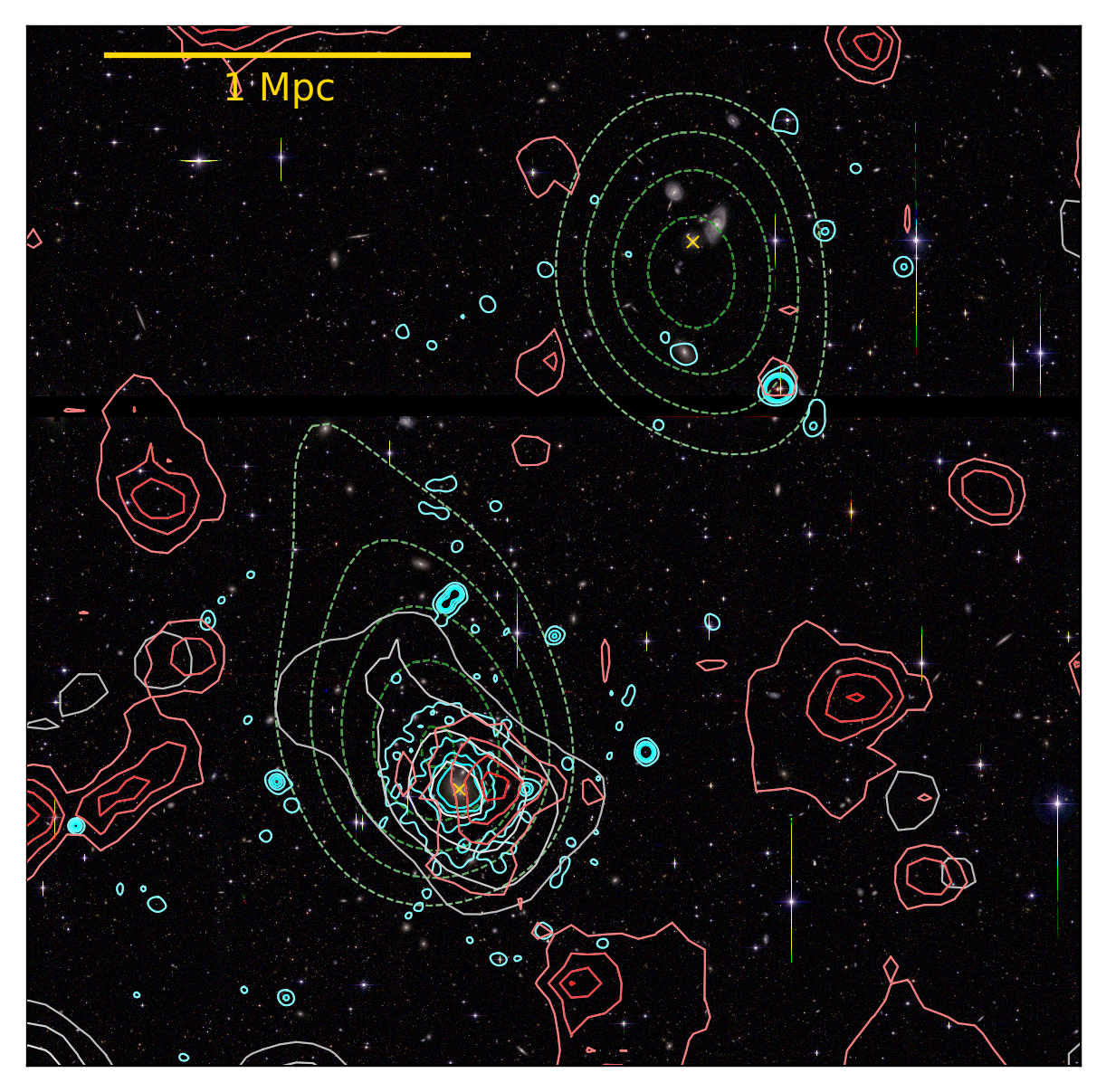}
    \caption{Reconstructed 2D mass map is shown in white contours, ranging from $1.2\,\sigma$ to $2.64\,\sigma$. X-ray emission from XMM-Newton is illustrated with cyan contours, ranging from 5$\sigma$ to 25$\,\sigma$. The green dashed contours indicate the luminosity map. The background shows the rgb image coadded from the $i$, $r$ and $g$ bands. The BCG and the brightest galaxy of the northwest region are marked with yellow crosses. Red contours are the results given by \texttt{Glimpse} with $\lambda=3.5$}.
    \label{fig:mass_map}
\end{figure}

% try bootstrap ?
Using the shear catalog described in Section \ref{subsect:sourcc_selection}, we performed mass reconstruction using two methods. For the standard Fourier inversion method \citep[hereafter KS93]{ks93}, we binned the shear map into a $90 \times 90$ pixel grid. The shear can be transformed into E- and B-mode convergence fields $\kappa_E$ and $\kappa_B$, respectively, as:
\begin{equation}
    \widetilde{\kappa}_E = ((k^2_1 - k^2_2) \widetilde{\gamma}_1 + 2 k_1 k_2 \widetilde{\gamma}_2) / (k^2_1 + k^2_2),
\end{equation}
and
\begin{equation}
    \widetilde{\kappa}_B = ((k^2_1 - k^2_2) \widetilde{\gamma}_2 - 2 k_1 k_2 \widetilde{\gamma}_1) / (k^2_1 + k^2_2),
\end{equation}
where the tildes denote Fourier transforms, and $k$ is the wave number. The obtained convergence map was then smoothed using an aperture mass filter from \cite{Schneider1998} with a radius of $13.3^\prime$ (20 pixels) to obtain the final mass map. In addition, we applied \texttt{Glimpse} \citep{Lanusse2016_glimpse}, which employs sparse regularization in the wavelet domain to recover convergence fields directly from irregularly sampled data without the need for prior binning.

Figure \ref{fig:mass_map} presents the reconstructed mass maps. The KS93 results show that the mass is predominantly concentrated in the southeast, coincident with the BCG position and consistent with the X-ray surface brightness. Notably, no significant over-density is visually detected in the northwestern region corresponding to the secondary optical peak. Additionally, although the \texttt{Glimpse} result is noisy, the mass distribution near BCG and NW optical peak is in agreement with KS93. 

\begin{figure}
    \centering
    \includegraphics[width=\linewidth]{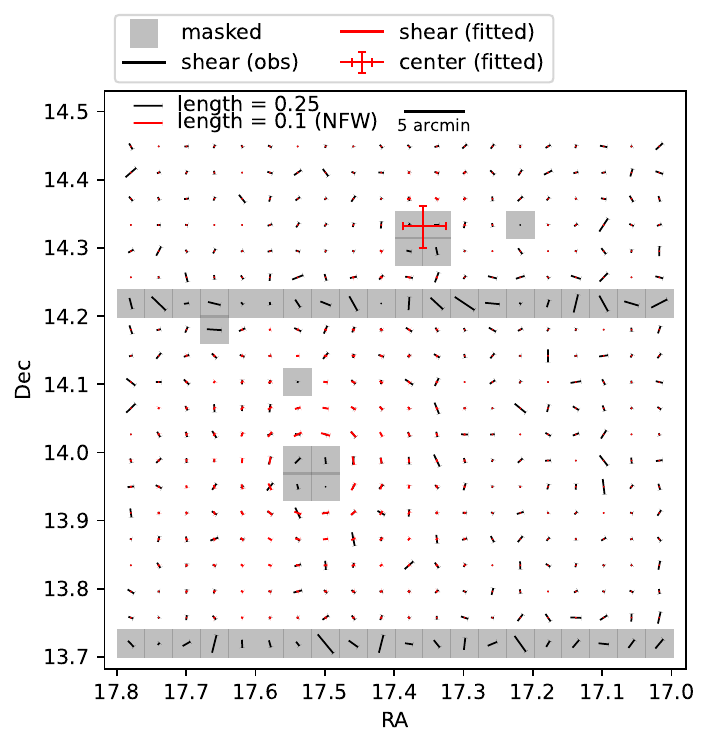}
    \caption{2D shear vector map. The black segments represent the binned observed shear, while the red segments show the best-fit model shear derived from the linear superposition of two NFW profiles. Grey regions indicate pixels masked due to low source galaxy density ($n_\mathrm{gal}<50$). The red cross marks the best-fit position of the northwestern (NW) component with 1 $\sigma$ uncertainty.}
    \label{fig:2dresult}
\end{figure}

To quantitatively assess the contribution of the northwestern substructure, we performed a parametric 2D shear fitting analysis assuming a dual halo model. The source redshift distribution is estimated using a subset of the COSMOS catalog \citep{cosmos2025}. Since CFHT $r$-band is not available in this catalog, we adopted the HSC $r$-band magnitude as a proxy. We selected a reference sample by applying the same magnitude cuts $21<m_{r,\text{HSC}}<26$ and a size cut of half light radius $> 0.5 \,\text{arcsec}$. While this size cut is stricter than the one applied to the shear catalog, it ensures that the reference redshift distribution is derived from well-resolved sources with reliable photometric redshifts, serving as a robust proxy for the lensing source population dominated by \texttt{lensfit} weighted galaxies.

As Figure \ref{fig:2dresult} shows, we selected a $0.8^\circ \times 0.8^\circ$ sub-field centered on the luminosity weighted centroid of the system and binned the shear data into a $20 \times 20$ grid to reduce shape noise. The mass distribution was modeled as the sum of two Navarro-Frenk-White (NFW) profiles \citep{NFW1997} halos at $z_\mathrm{cl}=0.05828$. The main halo was fixed at the BCG position, while the northwestern halo was initialized at the position of its brightest galaxy ($\mathrm{ra=01h09m25.92s}$, $\mathrm{dec=14^{\circ}19'55.20''}$) and allowed to vary.

Following \cite{Du2015}, the model shear at each pixel is the linear superposition of the shear fields generated by the two NFW profiles. To mitigate the degeneracy between mass and concentration given the limited signal-to-noise ratio, we did not fit the concentration as a free parameter. Instead, the concentration $c_{200}$ was directly determined from the mass $M_{200}$ using the mass-concentration relation of \cite{xu2021mc}. Consequently, only the masses of the two halos (and the position of the northwestern halo) were treated as free parameters. To ensure stability, we masked pixels containing fewer than 50 source galaxies, as well as the central $2\times2$ pixels of each NFW halo.

The fitting was performed with mass bounds of $12.5<\log(M_{200}/M_\odot)<15.5$ and a positional constraint of $0.1^\circ$ in both $\Delta\mathrm{RA}$ and $\Delta\mathrm{Dec}$ for the NW halo. We first used the Differential Evolution algorithm with Latin Hypercube Sampling to locate a maximum likelihood solution, which was then used to initialize the MCMC walkers. The 2D analysis yields $\log(M_{200}/M_\odot)=13.98^{+0.30}_{-0.50}$ for the SE main cluster and $\log(M_{200}/M_\odot)=13.07^{+0.50}_{-0.39}$ for the NW component. Within this model, the central NW mass estimate corresponds to a group scale system and is approximately an order of magnitude below the SE estimate. However, the broad uncertainty on the NW mass limits the precision of this comparison, and the fitted value does not by itself establish the presence of a distinct NW halo. A low mass group remains compatible with both the shear constraints and the absence of a significant NW peak in the reconstructed mass maps.

\begin{figure}
    \centering
    \includegraphics[width=\columnwidth]{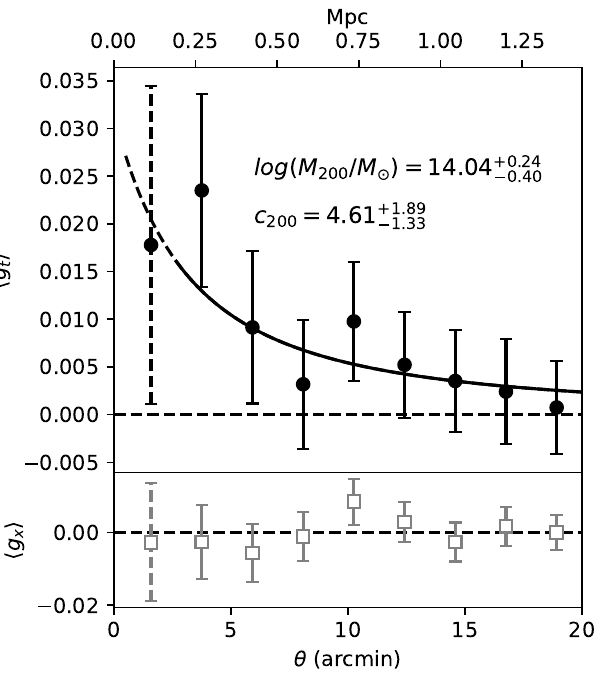}
    \caption{Radial dependence of the tangential (black dots) and cross components (grey hollow squares), the best-fit NFW profile is shown in black line. The inner region ($0.5' \text{--} 2.6'$), indicated by the dashed curve and dashed error bars, was not included in the fit and is plotted for display only.}
    \label{fig:gamma_t}
\end{figure}

% tangential shear profile
Given the dominance of the main halo, we proceed to characterize the cluster properties using a 1D azimuthally averaged tangential shear profile centered on the BCG. We decomposed the shear components into tangential ($\gamma_t$) and cross ($\gamma_\times$) components:
\begin{equation}
    \gamma_t = - \gamma_1 \cos(2\Phi) - \gamma_2 \sin(2\Phi),
\end{equation}
and 
\begin{equation}
    \gamma_{\times} = -\gamma_1 \sin(2\Phi) + \gamma_2 \cos(2\Phi)
\end{equation}
where $\Phi$ is the position angle with respect to the selected center. $\gamma_\times$ can provide a diagnostic test for systematic error, which should be zero for an ideal shear profile. We computed the azimuthally averaged shear profile by binning galaxies according to their projected radial distances from the cluster center. The bins range from 2.6 to 20 $\mathrm{arcmin}$ (corresponding to 0.2 -1.5 $\mathrm{Mpc}$ on the lens plane). The mean tangential shear within each bin is weighted by the \texttt{lensfit} weights. 

The observed profile is modeled using a single NFW halo centered on the BCG. Parameter estimation is also performed with MCMC. In contrast to the 2D analysis, we incorporated the full source redshift distribution to improve accuracy. Instead of adopting a single mean redshift, we integrated over the redshift distribution of the selected COSMOS source sample to compute the theoretical shear for each bin. Furthermore, we treated the concentration parameter $c_{200}$ as a free parameter. To mitigate degeneracy while allowing for intrinsic scatter, we employed the mass-concentration relation from \cite{xu2021mc} as a prior (rather than a deterministic constraint).Our fitting results and the tangential shear profile are shown in Figure \ref{fig:gamma_t}. This analysis yields a mass of $\log (M_{200}/M_\odot) = 14.04_{-0.40}^{+0.24}$ and a concentration of $c_{200} = 4.61_{-1.33}^{+1.89}$.

\subsection{Discussion}

Our multi-wavelength analysis reveals differences between the optical distribution of the Y12 selected system and the locations of its detected mass and hot-gas concentrations. The 71 galaxies assigned to this system by Y12 form two spatially separated concentrations in the SE and NW, with comparable numbers of galaxies. The primary weak lensing mass overdensity is located near the SE X-ray emission centroid, whereas no significant weak lensing peak or extended X-ray counterpart is detected at the NW optical center. These observations motivate a distinction between the full catalog selected galaxy sample and the SE-centered main cluster when interpreting the apparent rotation signal.

By fitting a single NFW profile to the tangential shear signal of the SE main cluster, we measure $\log(M_{200}/M_\odot)=14.04^{+0.24}_{-0.40}$. Although lower than the Y12 mass estimate, this result is statistically consistent with the mass inferred from the $M_{200}-T_{\mathrm{X}}$ relation, $\log(M_{200}/M_\odot)=14.202^{+0.061}_{-0.058}$, within the quoted measurement uncertainties. Together with the spatial correspondence between the primary weak lensing overdensity and the extended X-ray emission, these results identify the SE region as the main detected cluster scale mass concentration. Fitting two NFW halos to the two dimensional shear field yields a NW mass of $\log(M_{200}/M_\odot)=13.07^{+0.50}_{-0.39}$. Within this model, the central estimate corresponds to a group scale system and is approximately an order of magnitude below the mass scale inferred for the SE main cluster. However, the uncertainty in the NW mass and the limited sensitivity of the weak lensing reconstruction to low mass halos preclude a precise determination of the mass contrast from these central estimates alone. The absence of a significant NW peak in the reconstructed mass map therefore does not exclude an associated low mass halo. Likewise, the non-detection of extended X-ray emission at the NW optical center, particularly given the short usable exposure of the NW observation, does not rule out a low mass galaxy group.

% explain XCS J0107.3+1416 
Additionally, the nearby X-ray selected cluster XCS J0107.3+1416, associated with SDSS-C4 2049, lies approximately $43^\prime$ from the SE BCG and $31^\prime$ from the NW optical pezheak. It is therefore spatially distinct from the NW concentration and does not provide an X-ray counterpart at the NW optical center. In Y12, this system is identified as group 550 in modelA and group 808 in modelC. Our spectroscopic checks using members of these two catalog identifications place XCS at $z\simeq0.075$, whereas the 31 independently confirmed NW galaxies have a mean redshift of $z=0.060739$. The corresponding redshift space velocity separation is approximately $4000$--$4100~\mathrm{km\,s^{-1}}$, depending on the Y12 catalog version. None of these 31 NW galaxies lies within the $\pm3\sigma_v$ interval of either XCS spectroscopic sample. These results argue against contamination of the spectroscopically confirmed NW population by XCS members, while leaving the nature of the dynamical association between NW and the SE main cluster unresolved.

Although the center of XCS lies outside our weak lensing field of view, its halo can contribute to the observed shear field. We modeled XCS as an NFW halo at $z=0.07547$ and evaluated its contribution at the positions of the selected lensfit sources. Using the Y12 catalog halo masses as $M_{200\mathrm{m}}$, we converted the luminosity and stellar mass based proxies from modelA and modelC to reference $M_{200\mathrm{c}}$ values of approximately $(1.4$--$2.0)\times10^{14}\,M_\odot$ in our adopted cosmology. For these reference models, the predicted reduced shear amplitude is $|g_{\mathrm{XCS}}|\simeq10^{-3}$ at the SE center and $\simeq2\times10^{-3}$ at the NW center. Although the modeled shear increases toward the field boundary closest to XCS, its weighted contribution to the azimuthally averaged tangential shear around SE is less than 3\% of the corresponding measurement uncertainty in every fitted radial bin. Conditional maximum a posteriori (MAP) comparisons using the same lensfit inputs with and without a fixed XCS halo yield a decrease of approximately 0.4\%--0.6\% in the SE single-halo mass for the nominal reference models. The SE mass changes remain below 1\% across the additional XCS mass, concentration, centering, and outer-profile configurations tested. These comparisons quantify differential sensitivity to XCS and do not constitute a recalculation of the reported mass posteriors. The NW response depends more strongly on the treatment of constant residual shear components, its fitted center, and the assumed outer extent of the XCS halo. Consequently, the tests do not establish a unique correction to the NW mass. We use these calculations only as an external-halo sensitivity test.

The weak lensing mass of the SE main cluster implies $r_{200}\approx1~\mathrm{Mpc}$, corresponding to approximately $14^\prime$ at $z=0.058$. The NW optical concentration lies at a projected separation of $22^\prime$, or approximately $1.6r_{200}$, from the SE BCG. Although this places the NW center outside the estimated $r_{200}$ of the main cluster, it does not establish that the two populations are physically unrelated. The NW galaxies have bluer colors (Figure \ref{fig:color}) and a lower velocity dispersion than the SE population. These properties are compatible with a filamentary environment, but do not uniquely distinguish it from a low mass group. We therefore consider both a low mass group that may be undergoing infall and a filamentary population projected near the main cluster as viable interpretations. These possibilities are not mutually exclusive, since an infalling group may be embedded in a filament. The present observations do not establish whether the NW population constitutes a distinct gravitationally bound halo or whether it is dynamically bound to the SE cluster.

The full Y12 selected sample exhibits significant spatial--kinematic substructure, and the fitted projected rotation axis approximately separates the SE and NW concentrations. As discussed by MP17, projected groups and coherent substructure motions can mimic a cluster rotation signature. In particular, galaxy populations that are separated on the sky and have different mean redshifts can produce a rotation like pattern when analyzed together. Given the comparable numbers of SE and NW galaxies, the NW population can substantially influence this statistic even if its associated halo is much less massive than the SE cluster. The observed mean redshift difference may reflect line-of-sight distance separation, relative peculiar motions, or both; it does not by itself establish a physical association between the two populations. Together with the localization of the main detected mass and hot gas in SE, these findings favor interpreting the rotation like signal as arising from the combined analysis of two spatially separated populations with different mean redshifts.

\section{SUMMARY}  \label{sec:sum}

We investigated the candidate rotating cluster RXCJ0110.0+1358, also known as Abell 152, using galaxy kinematics, weak lensing, and X-ray observations. The 71 galaxies assigned to this system by Y12 form two spatial concentrations in the SE and NW. Applying the MP17 method to the combined sample reveals a rotation like signal, with the best fitting projected rotation axis effectively separating the two concentrations. The $\delta$-test also detects significant spatially correlated kinematic substructure: none of 10,000 random permutations of the galaxy velocities at fixed positions yields a cumulative statistic $\Delta$ as large as the observed value.

The NW galaxies are bluer and kinematically colder than the SE population, with velocity dispersions of $\sigma_{\mathrm{NW}}=326\,(331)~\mathrm{km\,s^{-1}}$ and $\sigma_{\mathrm{SE}}=548\,(548)~\mathrm{km\,s^{-1}}$, respectively; values in parentheses refer to the 56-galaxy spectroscopic subsample. 

Using new deep $r$-band imaging from CFHT MegaCam, we reconstructed the weak lensing mass distribution and identified its primary overdensity near the SE BCG and extended X-ray emission. A single NFW fit to the SE tangential shear profile gives $\log(M_{200}/M_\odot)=14.04^{+0.24}_{-0.40}$, consistent within the quoted measurement uncertainties with the X-ray temperature based estimate of $\log(M_{200}/M_\odot)=14.202^{+0.061}_{-0.058}$. The weak lensing central estimate is lower than the Y12 catalog value of $\log(M_{200}/M_\odot)=14.43$. The spatial correspondence between the detected mass and hot gas identifies SE as the main detected cluster scale mass concentration.

No significant weak lensing peak or extended X-ray counterpart is detected at the NW optical center. However, the limited weak lensing sensitivity to low mass halos and the short usable NW X-ray exposure do not exclude a low mass group. A potentially infalling group and a filamentary population remain viable interpretations, including the possibility of a group embedded in a filament. Given the comparable numbers of SE and NW galaxies, the NW population can substantially influence the combined velocity statistic even if its associated halo is much less massive than the SE cluster. Taken together, our multi-wavelength and kinematic results favor an interpretation in which the SE and NW galaxy populations, separated on the sky and differing in mean redshift, produce a rotation like signal when analyzed together.

\begin{acknowledgments}
We would like to thank the Canada-France-Hawaii Telescope (CFHT) Operations, Instrumentation, and Software Groups for their contributions and diligence in maintaining observatory operations; the CFHT Astronomy Group for their observation coordination and data acquisition efforts; and the CFHT Finance \& Administration Group for their contributions to the management and administration of the observatory.

L.P. F. acknowledges the supports from NSFC No. 12541302, the Innovation Program of Shanghai Municipal Education Commission (Grant No. 2025GDZKZD04),and China Manned Space Project with Grant No. CMS-CSST-2025-A05. W.D. acknowledges the support from the NSFC under grant No. 12541302 and the China Manned Spaced (CMS) program with grant Nos. CMS-CSST-2025-A02, and CMS-CSST-2025-A03. R. W. and D. T. were supported by a NASA/ADAP grant to Michigan State University  (NASA-80NSSC22K0476). W.L. and B.L. acknowledge the support from the National Key Research and Development Program of China (2023YFA1608100). B.L. acknowledges the Double-Innovation Doctor Program of Jiangsu Province (No. JSSCBS0216). WL acknowledges the support from the National Key R\&D Program of China (2021YFC2203100, 2018YFA0404503), NSFC(NO. 12573005, 11833005, 12192224, 11890693)
    
\end{acknowledgments}


\vspace{5mm}
\facilities{CFHT (MegaCam)}


\software{ccdproc \citep{ccdproc},
          %EAZY \citep{brammer2008eazy},
          lensfit \citep{miller2013lensfit},
          photutils \citep{photutils},
          SAS \citep{Gabriel2004},
          SCAMP \citep{Bertin2006}, 
          Source Extractor \citep{Bertin1996}, 
          SWarp \citep{Bertin2010},
          XSPEC \citep{Arnaud1996}
          }



% \appendix

\bibliography{reference}{}
\bibliographystyle{aasjournal}



\end{document}